\documentclass[twoside,twocolumn,a4paper,9pt]{article}
\usepackage{extsizes}
\usepackage[super,sort&compress,comma]{natbib} 
\usepackage[version=3]{mhchem}
\usepackage[left=1.5cm, right=1.5cm, top=1.785cm, bottom=2.0cm]{geometry}
\usepackage{balance}
\usepackage{times,mathptmx}
\usepackage{sectsty}
\usepackage{graphicx} 
\usepackage{lastpage}
\usepackage[format=plain,justification=justified,singlelinecheck=false,font={stretch=1.125,small,sf},labelfont=bf,labelsep=space]{caption}
\usepackage{float}
\usepackage{fancyhdr}
\usepackage{fnpos}
\usepackage[english]{babel}
\usepackage{array}
\usepackage{droidsans}
\usepackage{charter}
\usepackage[T1]{fontenc}
\usepackage[usenames,dvipsnames]{xcolor}
\usepackage{setspace}
\usepackage[compact]{titlesec}
\usepackage{amsthm}
\usepackage{bm}
\usepackage{hyperref}
\usepackage[retainorgcmds]{IEEEtrantools}
\usepackage{epstopdf}

\definecolor{cream}{RGB}{222,217,201}

\begin{document}

\pagestyle{fancy}
\thispagestyle{plain}


\makeFNbottom
\makeatletter
\renewcommand\LARGE{\@setfontsize\LARGE{15pt}{17}}
\renewcommand\Large{\@setfontsize\Large{12pt}{14}}
\renewcommand\large{\@setfontsize\large{10pt}{12}}
\renewcommand\footnotesize{\@setfontsize\footnotesize{7pt}{10}}
\makeatother

\renewcommand{\thefootnote}{\fnsymbol{footnote}}
\renewcommand\footnoterule{\vspace*{1pt}%
\color{cream}\hrule width 3.5in height 0.4pt \color{black}\vspace*{5pt}} 
\setcounter{secnumdepth}{5}

\makeatletter 
\renewcommand\@biblabel[1]{#1}            
\renewcommand\@makefntext[1]%
{\noindent\makebox[0pt][r]{\@thefnmark\,}#1}
\makeatother 
\renewcommand{\figurename}{\small{Fig.}~}
\sectionfont{\sffamily\Large}
\subsectionfont{\normalsize}
\subsubsectionfont{\bf}
\setstretch{1.125} 
\setlength{\skip\footins}{0.8cm}
\setlength{\footnotesep}{0.25cm}
\setlength{\jot}{10pt}
\titlespacing*{\section}{0pt}{4pt}{4pt}
\titlespacing*{\subsection}{0pt}{15pt}{1pt}

\renewcommand{\headrulewidth}{0pt} 
\renewcommand{\footrulewidth}{0pt}
\setlength{\arrayrulewidth}{1pt}
\setlength{\columnsep}{6.5mm}
\setlength\bibsep{1pt}

\makeatletter 
\newlength{\figrulesep} 
\setlength{\figrulesep}{0.5\textfloatsep} 

\newcommand{\topfigrule}{\vspace*{-1pt}%
\noindent{\color{cream}\rule[-\figrulesep]{\columnwidth}{1.5pt}} }

\newcommand{\botfigrule}{\vspace*{-2pt}%
\noindent{\color{cream}\rule[\figrulesep]{\columnwidth}{1.5pt}} }

\newcommand{\dblfigrule}{\vspace*{-1pt}%
\noindent{\color{cream}\rule[-\figrulesep]{\textwidth}{1.5pt}} }

\makeatother


\twocolumn[
  \begin{@twocolumnfalse}
\vspace{3cm}
\sffamily
\begin{tabular}{m{4.5cm} p{13.5cm} }

 & \noindent\LARGE{\textbf{Prescribing Patterns  in Growing Tubular Soft Matter by Initial Residual Stress}} \\
\vspace{0.3cm} & \vspace{0.3cm} \\

 & \noindent\large{Yangkun Du,\textit{$^{a,b}$}  Chaofeng L\"{u},\textit{$^{c,d,e,\ddag}$}, Congshan Liu,\textit{$^{c}$} Zilong Han,\textit{$^{a}$} Jian Li,\textit{$^{a}$} Weiqiu Chen,\textit{$^{a,d,e}$} Shaoxing Qu\textit{$^{a,d,e}$}, and Michel Destrade\textit{$^{b,a}$}} \\

& \noindent\normalsize{Initial residual stress is omnipresent in biological tissues and soft matter, and can affect  growth-induced pattern selection significantly. 
Here we demonstrate this effect experimentally by letting gel tubes grow in the presence or absence of initial residual stress and by observing different growth pattern evolutions.  
These experiments motivate us to model the mechanisms at play when a growing bilayer tubular organ spontaneously displays buckling patterns on its inner surface.  
We demonstrate that not only differential growth, geometry and elasticity, but also initial residual stress distribution, exert a notable influence on these pattern phenomena. 
Prescribing an initial residual stress may offer an alternative or a more effective way to implement a pattern selection for growable bio-tissues or soft matter. The results  also show promise for the design of 4D bio-mimic printing protocols or for controlling hydrogel actuators.} \\

\end{tabular}

 \end{@twocolumnfalse} \vspace{0.6cm}

  ]

\renewcommand*\rmdefault{bch}\normalfont\upshape
\rmfamily
\section*{}
\vspace{-1cm}


\footnotetext{\textit{$^{a}$~Department of Engineering Mechanics, Zhejiang University, Hangzhou, P. R. China;}}
\footnotetext{\textit{$^{b}$~School of Mathematics, Statistics and Applied Mathematics, NUI Galway, Galway, Ireland;}}
\footnotetext{\textit{$^{c}$~Department of Civil Engineering, Zhejiang University,  Hangzhou 310058, P.R. China; }}
\footnotetext{\textit{$^{d}$~Key Lab of Soft Machines and Smart Devices of Zhejiang Province, Zhejiang University, Hangzhou, P.R. China;}}
\footnotetext{\textit{$^{e}$~Soft Matter Research Center, Zhejiang University, Hangzhou 310027, P. R. China.}}

\footnotetext{\textit{$^{\ddag}$~Fax: 86-571-8820-8685; Tel: 86-571-8820-8473; E-mail: lucf@zju.edu.cn}}

%


%
%
%

\section{Introduction}
\begin{figure}
	\centering
	\includegraphics[width=0.41\textwidth]{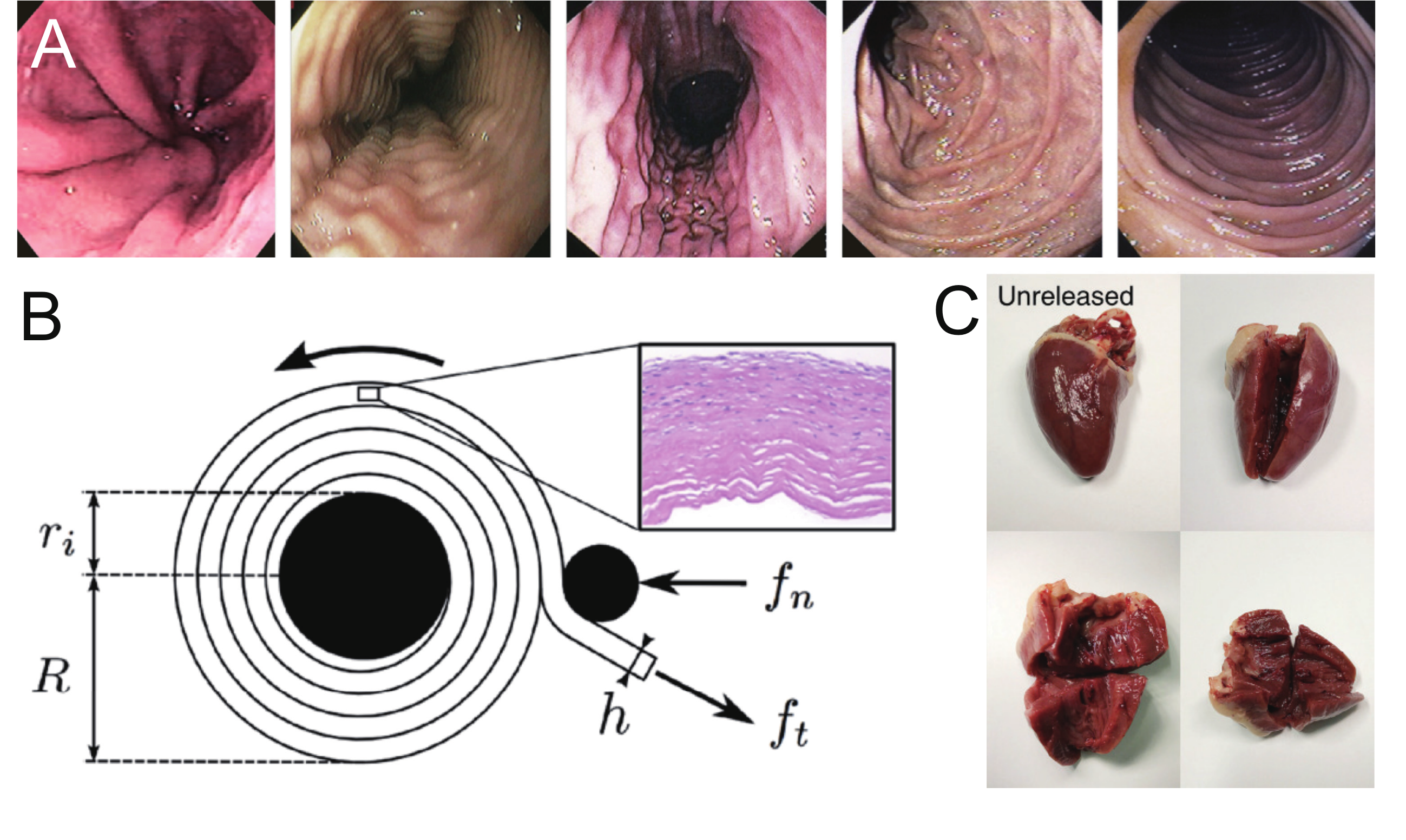}
\caption{
(A) Different morphological wrinkling patterns along the human intestine \cite{wilcox2012};
(B) A protocol for printing non-Euclidean solids, similar to 3D-printing bio-tissues in bioengineering \cite{zurlo2017}; 
(C) Residual stresses can be released in a duck heart by cutting it in different directions \cite{Du2018}.
}
	\label{Liver}
\end{figure}

Pattern creation in soft solids is a common phenomenon in Nature and is now being promoted in biomedical and industrial applications. 
In biological systems, some specific patterns are used to maintain essential bio-functions such as the wrinkles found in the intestine (Fig.~\ref{Liver}A), useful for digesting, and the interconnected creases of the brain cortex, associated with intelligence development \cite{Balbi2015, altman2005human}. 
Other patterns are used to transmit pathological changes: hence, an abnormally wrinkled airway indicates asthmatic bronchiole, and frequent morphological changes of a tumor point to a pre-metastatic state \cite{Roussos:2011aa}. 
In other words, Nature and Evolution have mastered well how to control and select  optimal patterns for soft tissues. 
In turn, engineers try to mimic these processes to recreate and control ideal patterns in manufacturing. 
For example, self-assembly of a substrate-film structure for 3D micro-fabrication is achieved by local mismatch deformation between substrate and film: 
the resulting substrate curvature  underpins spontaneous micro- or nano-patterns and structures \citep{Yin2009, Chen2010, Chen2013}. 
Similarly, stimuli-responsive hydrogel actuators can be designed by prescribing inhomogeneous swelling  and made to wrinkle into different patterns.

Physically,  wrinkle/crease patterns  are created by mechanically-induced instability and post-buckling process. 
The special case of spontaneous instability and post-buckling in the absence of external loads can be explained by the presence of inhomogeneous \emph{residual stress fields}. 
These have been demonstrated experimentally for many bio-tissues such as skin, arteries, heart, brain, intestine, solid tumors, etc. \citep{Holzapfel2007, Holzapfel2010, Omens1990, Savin2011, Balbi2015, Stylianopoulos2012, Fernandez-Sanchez2015}, where they are required to ensure self-equilibrium, transfer bio-signals, or maintain some specific bio-functions \citep{Pan2016, Padera2004, Helmlinger1997}.
They can also endow  elastic materials with prescribed properties (Fig.~\ref{Liver}B).
A strong effort has been dedicated to model growth-induced residual stresses and the resulting pattern-generating instability \citep{ Cao2012, Li2011, Lu2016, Du2017, Balbi2015}.
 These papers used volume growth theory \cite{Rodriguez1994}, where growth is initiated from an initially stress-free configuration. 
In particular, the influence of differential volume growth and growth velocity on the creation of residual stress in bilayer cylinders was elucidated, resulting in a pattern selection protocol which can be tuned by changing  thickness and stiffnesses ratios \citep{Rodriguez1994, Goriely2005, Ciarletta2013, Balbi2015}.


%
%
%
However, most living tissues do not possess a stress-free initial configuration, as can be checked by cutting them in different directions: each cut releases some residual stress and consequently an infinite number of cuts are required to attain this hypothetical zero stress initial configuration (see the examples of a cut duck heart in Fig.~\ref{Liver}B or cut human arteries in \cite{Holzapfel2007}). 
In reality, long-term growth and remodelling processes and other bio-interactions are impossible to track and reconstruct for living matter. Hence,  the stress-free assumption for the initial state is too strong, and neglecting initial residual stress may affect the analysis of growth-induced residual stress and the resulting pattern selection, as we have proofed in our previous paper \cite{Du2019} on the growth human aorta with different initial residual stress.

In this current paper, we designed an experiment with swelling hydrogel and different residual stresses preset by shrink-fitting way to demonstrate the influence of initial residual stress on growth and the significance of our series works. 
Furthermore, by simplifying the initial radial stress as linear distribution with a magnitude factor instead of the specific and complex residual stress distribution, we are able to recover qualitatively all possible residual stress distributions.   Both the experimental and theoretical results have the consistent conclusions that the initial residual stress will not only affect the growth-induced residual stress and the pattern evolution but also do impact the onset of the critical buckling which is very important in practice. 

We structure this paper as follow. In Section 2, we show the experiment process and the pattern evolution phenomenons which assess the significance of the initial residuals on growth and morphologies. 
Section 3 is about the mathematical modelling on growth and instability analysis corresponding to the experiment.  Also, the typical residual stress distribution is provided based on self-equilibrium and boundary conditions. 
In section 4 and 5, we disclose some results and conclusions. 


\section{Experiments} 
Here, we designed an experiment to illustrate the influence of the initial residual stress on the growth-induced pattern evolutions. We use the shrink-fitting way to prescribe the initial residual stress into the bilayer structures which are made of the hydrogel inside and rubber outside. The hydrogel here is used to simulate the isotropic growth process. 
\begin{figure}[h]
	 \centering
	 \includegraphics[width=0.50\textwidth]{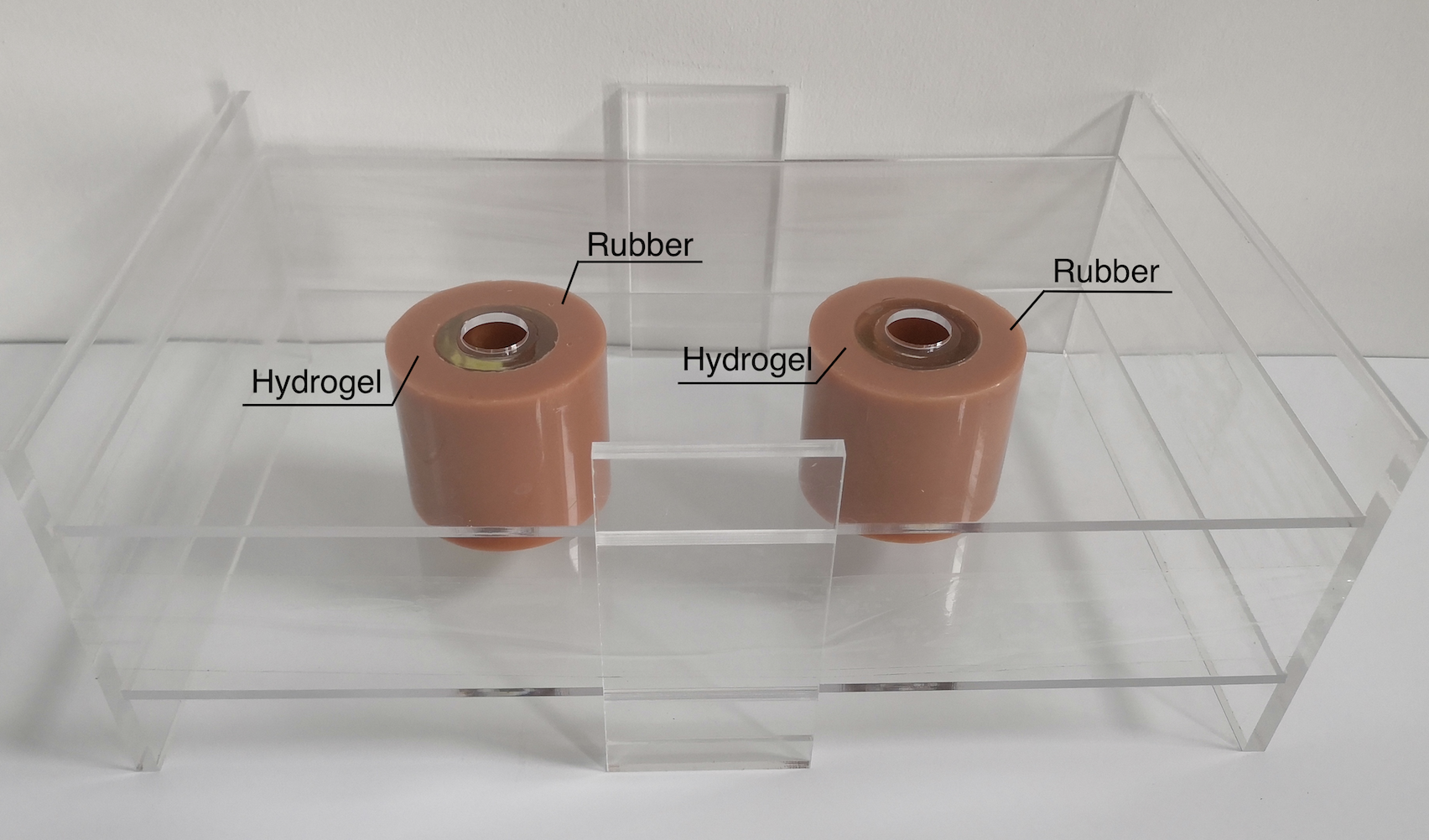}
	\caption{Experimental equipment for isotropic growth-induced instability of bilayer tubes, one with non-zero initial residual stress, one without.}
	\label{equipment}
\end{figure}

\begin{figure*}[h]
	\centering
	\includegraphics[width=\textwidth]{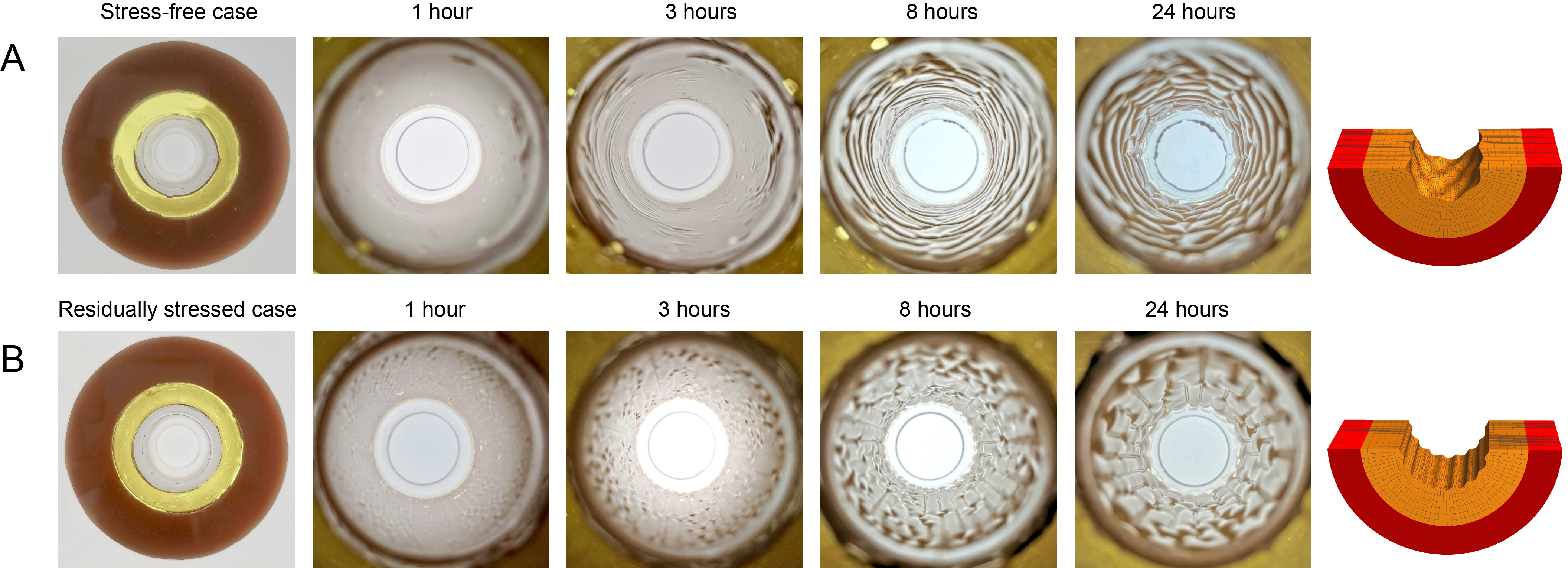}
	\caption{Growth-induced buckling of bilayer tubes (inner layer: swelling hydrogel, outer layer: inert rubber) with (A) stress-free and (B) residually stressed initial states. By  shrink-fitting, we created a non-zero initial residual stress in (B) (compressive hoop stress in the hydrogel, tensile hoop stress in the rubber). Otherwise, (A) and (B) have the same initial geometry and material parameters ($R_o/ R_s \simeq 1.8$, $R_s / R_i =1.5$,  $\mu^\text{inn} / \mu^\text{out} \simeq 0.2$). 
	Axial growth is prevented. Once submerged in water, (B) buckles earlier and in a different pattern (wrinkle-front aligned with the axis) than (A) (mixed axial/circumferential wrinkles).}
	\label{exp}
\end{figure*} 

\subsection{Protocol}
 
First, we prescribed an initial residual stress by setting an incompatible geometry for the two separate parts (hydrogel and rubber). 
We preset the outer radius of the inner layer hydrogel to be a little larger than the inner radius of the outer layer rubber. 
Then, by  shrink-fitting the hydrogel into the rubber, we created a compressive hoop stress in the hydrogel and a tensile hoop stress in the rubber. 
We measured the changes in radii due to the shrink-fitting and found them to be negligible (less than 1 mm).

For direct comparison with the case where there is no initial stress, we nested another hydrogel tube into another rubber tube, with the same geometry and materials as the first bilayer tube, but no incompatibility (the outer radius of the hydrogel was equal to the inner radius of the rubber). 

All tubes were 7 cm tall. The inner hydrogel tubes had inner radius $R_i = 30$ mm and outer radius $R_s = 45$ mm. 
The outer rubber tubes had inner radius $R_s = 45$ mm and outer radius $R_o = 80$ mm.  The elastic modulus of inert rubber is 448 kPa, while the elastic modulus of swelling hydrogel is 87 kPa. 

We placed the two bilayer tubes in an Acrylic frame, with fixed top and bottom plates, to constrain the deformation in the axial direction, see the experimental equipment shown in Figure \ref{equipment}. 
We pierced two circular holes on both the top and bottom plates, with centres on the axis of the tubes, for water to flow freely during the growth period.

Finally, the frame and tubes were placed in a big water tank.

\subsection{Materials}
Hydrogel: N,N’-methylenebis(Acrylamide) 98\% (MBAA, Lot \#146072) and Ammonium persulfate 98\% (APS, Lot \#248614) were purchased from Sigma-Aldrich; Acrylamide 99\% (AAm, Lot \#A108465) was purchased from Aladdin, Shanghai, China. All reagents were used as received.

Rubber: VytaFlex$^{TM}$ 30 was purchased from Smooth-On, Macungie, USA.


\subsection{Preparation}


Hydrogel: 
First, the AAm monomer is dissolved in distilled water to form a solution of concentration 4 mol/L. 
Then, to every 1 ml of the solution, 4 $\mu$l of a 0.1 mol/L of MBAA solution is added as the conventional cross-linker, and 20 $\mu$l of a 0.1 mol/L ammonium persulfate solution is added as the UV initiator. 
The resulting solution is degassed by N$_2$ for 1 hour and then poured into a mould made of laser-cut acrylic sheets. 
The mould and the solution are covered with the bottom of a Petri dish to prevent oxygen inhibition. 
The covered mould is then placed under a UV lamp and exposed to UV irradiation (wavelength = 365 nm). 
Finally, the hydrogel is taken out of the mould and washed with de-ionized water thoroughly to remove any unreacted monomers, and is kept at room temperature.

Rubber: See the fabrication instructions at \url{http://www.smooth-on.com.cn/uploadfile/2018/0510/20180510040704891.pdf}

\subsection{Experimental conclusions}

Fig.~\ref{exp} provides experimental proof of the impact of  initial residual stress on growth-induced pattern creation and evolution. 
The two bilayer cylinders have the same geometry and materials but different initial residual stresses. 
The prescribed initial residual stresses are stress-free  for one bilayer tube (Fig.~\ref{exp}A), and compressive hoop stress in hydrogel and tensile hoop stress in rubber for the other (Fig.~\ref{exp}B) 

The results show that by prescribing an initial residual stress,  we can bring forward the onset of instability and also create a pattern with straight circumferential folds along the axial direction, as opposed to the postponed, mixed circumferential and axial folds shown in Fig.~\ref{exp}A.
%

\section{Modeling}  
\subsection{Growth with initial residual stress}
Here we show that the so-called `modified multiplicative decomposition growth' (MMDG) model \cite{Du2018,Du2019} is  consistent with our experimental findings.

The MMDG framework relies on the concept of multiplicative decomposition, similar to the conventional volumetric growth model \cite{Rodriguez1994}. 
Its innovation lies in the introduction of an initial elastic deformation $\bm{F}_0$ which is used for releasing the initial residual stress to a virtual stress-free configuration. 
Thereafter, the independent unconstrained growth process $\bm{F}_g$ can take place freely, using two virtual stress-free and incompatible configurations. The elastic deformation $\bm{F}_e$ then makes the material compatible again.
Accordingly, the total growth process can be formulated by the deformation gradient
\begin{equation}
	\boldsymbol{F}=\boldsymbol{F}_e\boldsymbol{F}_g\boldsymbol{F}_0.
	\label{F}
\end{equation}

Here, we assume that the material in its virtual stress-free state is an incompressible neo-Hookean solid. 
Then the constitutive equation for isotropic growth is
\begin{equation}
	\boldsymbol{\sigma}=J_g^{-\frac{2}{3}} \left( \boldsymbol{F\tau F}^T+p_0\boldsymbol{FF}^T \right)-p\boldsymbol{I},
	\label{constitutive equation}
\end{equation}
where $p$, $p_0$ are the Lagrange multipliers in current and reference configurations, respectively, and $J_g=\det \boldsymbol{F}_g$ tracks local volume change. 
The initial stress $\boldsymbol{\tau}$ and the Cauchy stress  $\boldsymbol{\sigma}$ satisfy the equilibrium equations and boundary conditions in each configuration,
\begin{equation}
	\text{Div} \, \boldsymbol{\tau}=\boldsymbol{0},\quad \boldsymbol{\tau^}T \boldsymbol{N}=\boldsymbol{0},
	\quad \text{div} \, \boldsymbol{\sigma}=\boldsymbol{0},\quad\boldsymbol{\sigma}^T \boldsymbol{n}=\boldsymbol{0},
	\label{equilibrium equation}
\end{equation}
where $\boldsymbol n$, $\boldsymbol N$ are unit vectors normal to the boundary.

\begin{figure*}[h]
	{ \centering
	 \includegraphics[width=0.90\textwidth]{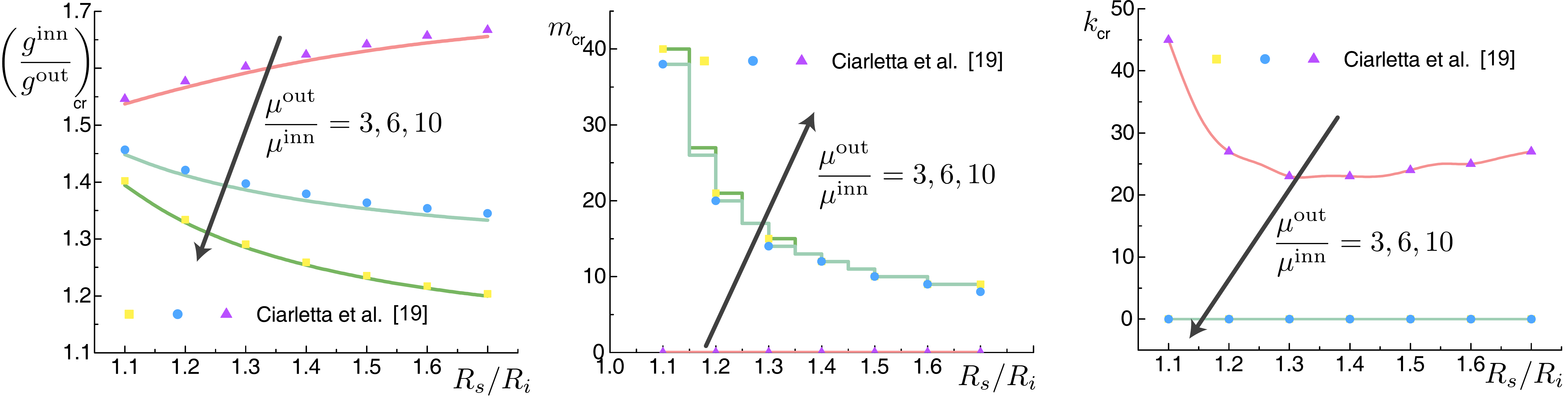}
	}
	\caption{Benchmark results of growth-induced instability without initial residual stress  \cite{Ciarletta2014}, where $R_o/R_s=1.8$, $R_o=1$, $g^{ad}=1$.}
	\label{PRL2014}
\end{figure*}

\subsection{Geometry and initial residual stress field}
For comparison with previous results on pattern selection in growing tubular soft solids \cite{Ciarletta2014}, we characterise our bilayer tube as being typical of bio-tissues. 
Hence, we take shear moduli in the ranges $\mu^\text{inn}=120-700$ Pa, $\mu^\text{out} = 1$ kPa (close to those of embryonic gastrointestinal tissue).  We take the the outer layer radius $R_\text{o}=1$, interface radius $R_\text{s}=R_\text{o}/1.8$ and inner radius $R_\text{i}$ as a geometric variable. 

To reproduce the initial residual stress in the experiments and in some bilayer tissues, we prescribe a linear variation of the radial initial residual stress, as
\begin{equation}
	\begin{split}
	&\tau_{RR}^\text{inn}=\frac{\alpha}{R_s-R_i} (R - R_i),\\[4pt]
	&\tau_{RR}^\text{out}=\frac{\alpha}{R_o-R_i} (R - R_s),\\
	\end{split}
	\label{IRS}
\end{equation}
which satisfies the required boundary conditions at the inner ($R_i$), interface ($R_s$) and outer ($R_o$) surfaces of the bilayer. 
We take the axial initial residual stress as $\tau_{ZZ}^{inn}=\tau_{ZZ}^{out}=0$. 
Then we obtain the distribution of the circumferential stress $\tau_{\Theta\Theta}$ by solving $\text{Div} \; \boldsymbol{\tau}= \boldsymbol 0$, the self-equilibrium equation for the initial residual stress.  

The initial residual stress field in Eq.\eqref{IRS} is compressive radial stress $\tau_{RR}$ and a hoop stress $\tau_{\theta\theta}$ which is compressive (tensile) in the inner (outer) layer, reproducing the qualitative characteristics of our experiments on hydrogel/rubber tubes. 
We pre-multiply this initial residual stress distribution by a magnitude factor $\alpha$, which we use to quantify the influence of the initial stress: $\alpha=0$ means a totally stress-free initial state, $\alpha>0$ means the inner layer is under compressive hoop initial stress (as in our experiments), $\alpha<0$ means the outer layer is hoop-compressed initially.
Finally we call $g^\text{inn}$, $g^\text{out}$ the volumetric growth factors in the inner and outer layers, respectively, and we solve the corresponding  buckling boundary value problem.

\subsection{Buckling analysis}


The onset of instability is analyzed by relying on linear incremental theory \cite{ogden1997non} and on our previous papers \cite{Du2018, Du2019}. 

In short, there exists a relation giving the increment $\boldsymbol{\dot{F}}$ of the displacement gradient $\boldsymbol F$ with respect to the reference configuration as $\boldsymbol{\dot{F}}=\boldsymbol{\dot{F}}_I\boldsymbol{F}$, where $\boldsymbol{\dot{F}}_I$ is the incremental displacement gradient with respect to the current configuration. Since the growth process is assumed to be independent of the stress and strain fields due to the infinitesimal and transient incremental deformation, we also have the relationship $\boldsymbol{\dot{F}_e} = \boldsymbol{\dot{F}}_I\boldsymbol{F_e}$ for the pure elastic gradient $\boldsymbol{F_e}$ and its increment.

By prescribing the incremental displacement field as 
\begin{equation}
	\boldsymbol{\dot{x}}=u\left(r,\theta, z\right) \boldsymbol{e}_r+v\left(r,\theta, z\right) \boldsymbol{e}_\theta+w\left(r,\theta, z\right) \boldsymbol{e}_z,
\end{equation}
we get the incremental displacement gradient tensor as
\begin{equation}
	\boldsymbol{\dot{F}}_I=\frac{\partial{\boldsymbol{\dot{x}}}}{\partial{\boldsymbol{x}}}=
	\begin{bmatrix}
		\dfrac{\partial{u}}{\partial{r}} &\dfrac{1}{r}\left(\dfrac{\partial u}{\partial \theta}-v\right) &\dfrac{\partial{u}}{\partial{r}} 	\\
		\dfrac{\partial{v}}{\partial{r}} &\dfrac{1}{r}\left(\dfrac{\partial v}{\partial \theta}+u\right) &\dfrac{\partial{v}}{\partial{r}} 	\\
		\dfrac{\partial{w}}{\partial{r}} &\dfrac{1}{r}\dfrac{\partial w}{\partial \theta} &\dfrac{\partial{w}}{\partial{z}}
	\end{bmatrix}.
\end{equation}
Then the incremental incompressibility  condition reads
\begin{equation}
	\mathrm{tr}\boldsymbol{\dot{F}_I}=\frac{\partial u}{\partial r}+\frac{1}{r}\left(\frac{\partial v}{\partial \theta}+u\right)+\frac{\partial{w}}{\partial{z}}=0.
	\label{incompressible condition}
\end{equation}

Now the incremental nominal stress $\boldsymbol{\dot{S}}$ in push-forward form has components 
\begin{equation}
	\dot{{S}}_{Iij}=A_{eijkl}^I \dot{{F}}_{Ilk}-\dot{p}\delta_{ij}+p\dot{{F}}_{Iij},
\end{equation}
where $A_{eijkl}^I=F_{ei\alpha} F_{ek\beta} \frac{\partial \psi}{\partial F_{ej\alpha}\partial{F_{el\beta}}}$ are the components of the instantaneous elasticity tensor, calculated by differentiating the strain energy density. 
Finally, the incremental stress equilibrium equations are:
\begin{equation}
	\text{div}\; \boldsymbol{\dot{S}}_I = \boldsymbol{0},\qquad\boldsymbol{\dot{S}}_I^T \boldsymbol{n}=\boldsymbol{0},
\end{equation}

The forms of incremental displacement  fields are specified as circumferential and axial sinusoidal wrinkles, in the form $\left[u,\dot{p} \right]=\left[U(r),Q(r) \right] \cos(m\theta) \cos(\kappa z)$, $v=V(r)\sin(m\theta) \cos(\kappa z)$, $w=W(r)\sin(m\theta) \cos(\kappa z)$, where $m$ and $\kappa= 2\pi n/L$ are the circumferential and axial wave numbers, respectively. 
Then the incremental nominal stress components are similar, as $ \dot{S}_{Irr}=\Sigma_{rr} \left(r\right) \cos \left(m\theta\right) \cos \left(\kappa z\right)$, $\dot{S}_{Ir\theta}=\Sigma_{r\theta} \left(r\right) \sin \left(m\theta\right) \cos \left(\kappa z\right)$, $\dot{S}_{Irz} =\Sigma_{rz} \left(r\right) \cos \left(m\theta\right) \sin \left(\kappa z\right)$. Finally, we arrive at the Stroh formulation of the incremental equations of equilibrium, as 
\begin{align} 
& \boldsymbol{\eta}(r)=[U(r), V(r), W(r), r\Sigma_{rr}(r), r\Sigma_{r\theta}(r), r\Sigma_{rz}(r)]^T,\notag \\[4pt]
& 	\frac{d}{dr} \boldsymbol{\eta}(r) =\frac{1}{r}
	\begin{bmatrix} 
	\boldsymbol{G}_1(r) & \boldsymbol{G}_2(r)\\
	\boldsymbol{G}_3(r) & \boldsymbol{G}_4(r)
	\end{bmatrix}
	\boldsymbol{\eta}(r),
	\label{Stroh1}
\end{align}
see  \cite{Du2018, Du2019} for details.

Finally, by iterating over the wrinkle numbers $m$ and $n$ for the numerical solution of Eq.~\eqref{Stroh1}, we obtain the critical value of each case which creates a buckling pattern for some given differential growth ratio. 
Here, we use the surface impedance method to integrate the Stroh formulation \citep{Biryukov1985350, DESTRADE20101212, DESTRADE20094322, SU2019}.

\section{Results}   
The numerical strategy is to find the smallest differential growth ratio $g^\text{inn}/g^\text{out}$ for which an incremental solution exists, for given wrinkle numbers $n$, $m$. 
Then after spanning all possible wrinkle numbers in the circumferential and axial directions, we keep the smallest ratio  $(g^\text{inn}/g^\text{out})_\text{cr}$ for the onset of buckling. If  the corresponding $n_\text{cr}$, $m_\text{cr}$ are both non-zero, then the wrinkling pattern is two-dimensional.

According to the expression of the initial residual stress in Eq.~\eqref{IRS}, we can recover instability results with a stress-free initial state by prescribing $\alpha=0$, and use the results as benchmark results, see Figure \ref{PRL2014}.
Hence we see that increasing the thickness or stiffness ratio of the outer to inner tubes will create fewer folds in the circumferential direction and more folds in the axial direction. 
A deep analysis of pattern selection by these factors has been conducted by \citet{Ciarletta2014}. 

In that case, which is not realistic for actual living matter (Fig.~\ref{Liver}B), pattern selection can only be tuned by the geometric and elastic parameters.
Starting then from an initially stressed state $\alpha \ne 0$, we find that pattern selection can be largely tuned or prescribed by the magnitude $\alpha$ of the initial stress $\boldsymbol{\tau}$.

\begin{figure}[h]
	 \centering
	\includegraphics[width=0.45\textwidth]{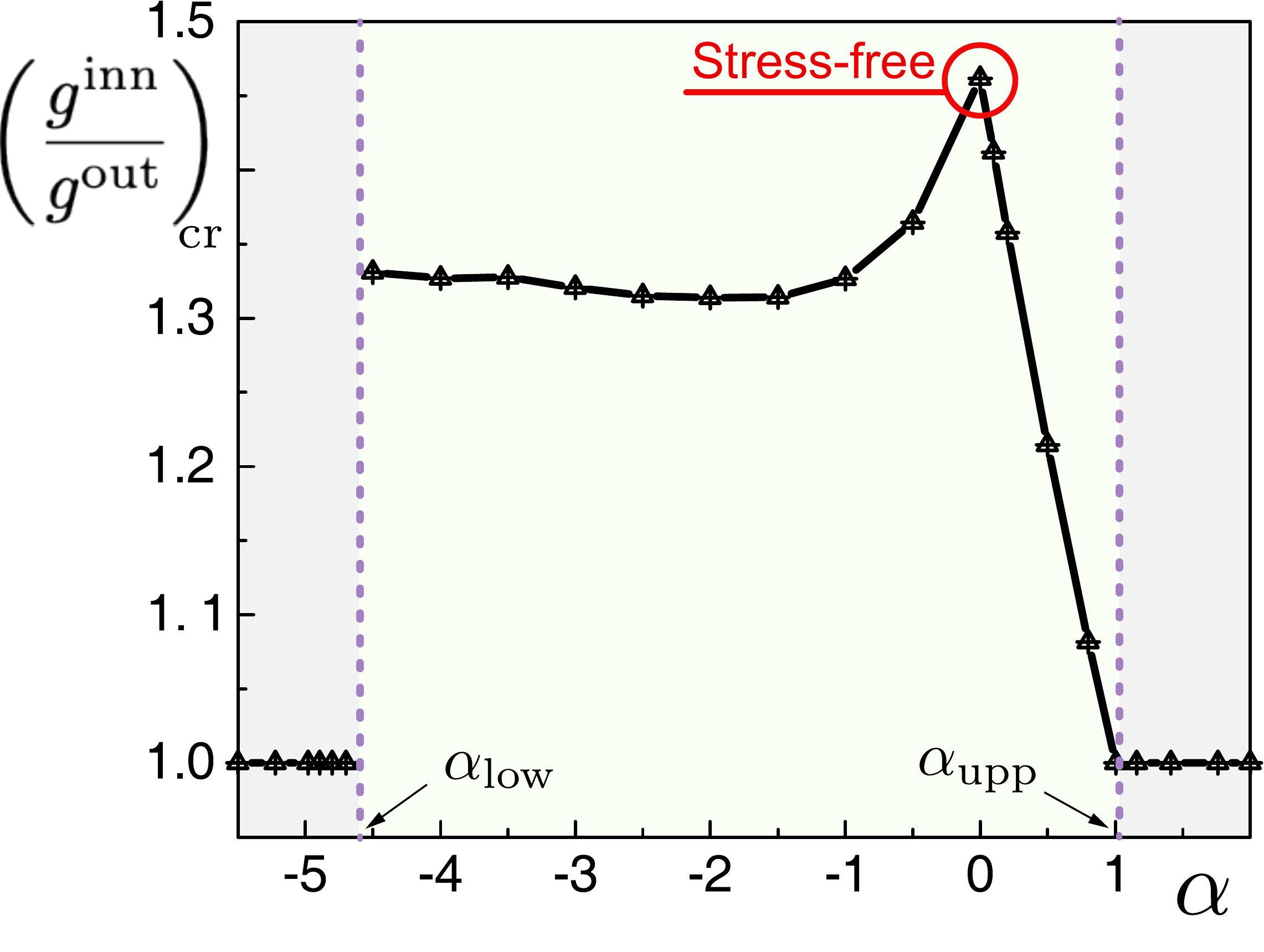}
	\caption{Variation of the critical differential growth ratio $(g^\text{inn}/g^\text{out})_\text{cr}$   with the level of initial stress $\alpha$, when the stiffness ratio is $\mu^\text{inn}/\mu^\text{out}=1/5$.}
	\label{GF_1}
\end{figure}

Generally,  soft tubular tissues are too soft to sustain increasing levels of initial residual stresses for long, especially compressive hoop stresses, which quickly induce buckling patterns in the absence of growth ($(g^\text{in}/g^\text{ad})_{cr}=1$) and external loads \cite{Ciarletta2016}. 
Here we find that this occurs when the amplitude of the initial stress is large enough. 
Hence, wrinkles appear on the inner face of the composite tube when $\alpha > \alpha_\text{upp} \simeq 1.003$, and on its outer face when $\alpha < \alpha_\text{low} \simeq -4.560$ (as the outer layer is then under large compressive hoop stress), see Fig.~\ref{GF_1}.


\begin{figure*}[h]
	\centering
	\includegraphics[width=.88\textwidth]{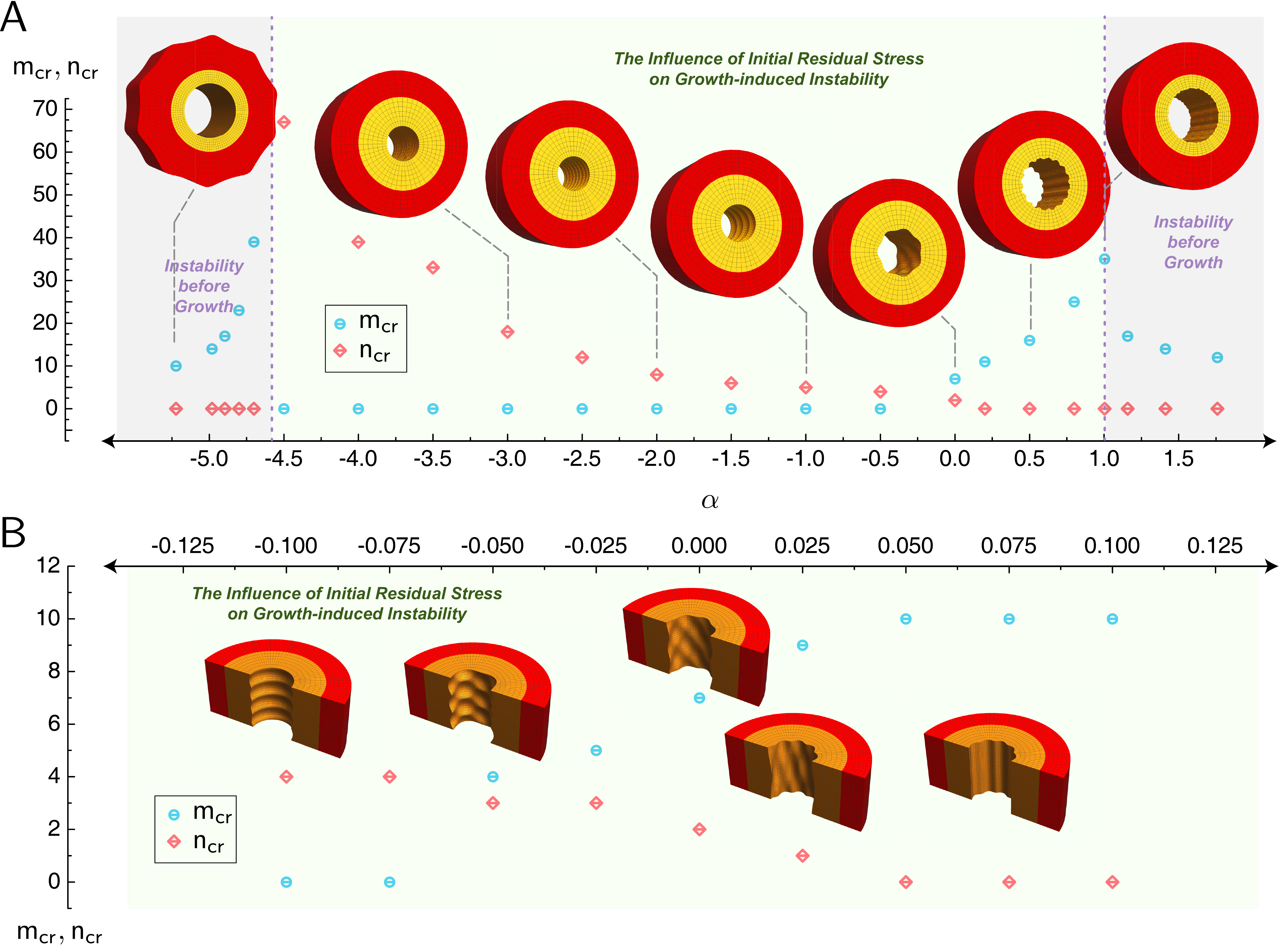}
	\caption{(A) Effect of the magnitude of the initial residual stresses $\alpha$ on  growth-induced pattern selection, when $\mu^\text{inn}/\mu^\text{out}=1/5$; (B) Zoom of the results in the range $-0.01<\alpha<0.01$ in (A).}
	\label{IIS_2}
\end{figure*}

Otherwise, when $\alpha_\text{low}<\alpha<\alpha_\text{upp}$,  instability is due to  combined high levels of initial stress and differential growth. 
Figs.~\ref{GF_1} and \ref{IIS_2} show the effect of the initial stress magnitude $\alpha$ on the evolution of growth-induced patterns. 
Fig.~\ref{GF_1} shows that with initial residual stress ($\alpha \ne 0$), the level of differential growth $(g^\text{inn}/g^\text{out})_\text{cr}$ required for buckling is reduced. 
It is worth noting here that this calculation result is consistent with our experimental finding in Fig.~\ref{exp} that the stress-free case occurs wrinkles earlier than the residual stress case.

Fig.~\ref{IIS_2} gives the numbers of circumferential and axial wrinkles. 
In the pure growth-induced case ($\alpha=0$), a 2D-pattern is predicted ($m_\text{cr}=7$, $n_\text{cr}=2$), consistent with our experiments (Fig.~\ref{exp}A). 
Then, increasing the initial circumferential compressive stress in the inner layer with $\alpha >0$ increases $m_\text{cr}$ and decreases $n_\text{cr}$. 
Quickly (see zoom of the $-0.1<\alpha<0.1$ range in Fig.~\ref{IIS_2}B), $n_\text{cr} =0$ and the wrinkles are aligned with the tube axis, consistent with our experiments (Fig.~\ref{exp}B).
Conversely, reversing the location of circumferential compressive and tensile  stresses by taking $\alpha<0$ leads to $m_\text{cr}=0$.  Fig.~\ref{IIS_2}B further displays the details and profiles of  possible patterns around $\alpha =0$, and shows how sensitive  they are to initial residual stress. 
Effectively, all the wrinkling scenarios encountered in the  intestine  (Fig.~\ref{Liver}A) can be captured by varying the magnitude and the sign of $\alpha$ only, while keeping the geometric and material parameters the same throughout.
The figures demonstrate how to obtain,  control and advance pattern creation by prescribing initial residual stress fields for given elasticities and geometries.
%
\\[8pt]
\section{Conclusions}  
We investigated the natural capability of biological tubular tissues to use initial residual stresses to control pattern creation, based on a recently developed growth model, and also mimicked the results experimentally by swelling hydrogel tubes. 
Initial stress is an effective and controllable factor for pattern selection beyond the geometric and elastic parameters highlighted in previous studies of pure growth \cite{Ciarletta2014}. 
We also showed that there is an effective range $\alpha_\text{low}<\alpha<\alpha_\text{upp}$ for the level of initial stress where patterns can be  prescribed on growable bio-tissues. 
We did not expand the instability analysis beyond the linearised buckling state, but it is worth noting that wrinkles are very stable for layered structures and give the number and wavelength of the eventual super-critical creases \cite{cao2011wrinkles}. 
They are also a way to measure the level of a known distribution of initial stress \cite{Ciarletta2016}. 

Our hope is that these results may provide an inspiring insight for directional bionic self-assembly or robot manufacturing by initial residual stress.  
\balance


\providecommand*{\mcitethebibliography}{\thebibliography}
\csname @ifundefined\endcsname{endmcitethebibliography}
{\let\endmcitethebibliography\endthebibliography}{}



\begin{mcitethebibliography}{36}
\providecommand*{\natexlab}[1]{#1}
\providecommand*{\mciteSetBstSublistMode}[1]{}
\providecommand*{\mciteSetBstMaxWidthForm}[2]{}
\providecommand*{\mciteBstWouldAddEndPuncttrue}
  {\def\EndOfBibitem{\unskip.}}
\providecommand*{\mciteBstWouldAddEndPunctfalse}
  {\let\EndOfBibitem\relax}
\providecommand*{\mciteSetBstMidEndSepPunct}[3]{}
\providecommand*{\mciteSetBstSublistLabelBeginEnd}[3]{}
\providecommand*{\EndOfBibitem}{}
\mciteSetBstSublistMode{f}
\mciteSetBstMaxWidthForm{subitem}
{(\emph{\alph{mcitesubitemcount}})}
\mciteSetBstSublistLabelBeginEnd{\mcitemaxwidthsubitemform\space}
{\relax}{\relax}

\bibitem[Balbi \emph{et~al.}(2015)Balbi, Kuhl, and Ciarletta]{Balbi2015}
V.~Balbi, E.~Kuhl and P.~Ciarletta, \emph{J. Mech. Phys. Solids}, 2015,
  \textbf{78}, 493--510\relax
\mciteBstWouldAddEndPuncttrue
\mciteSetBstMidEndSepPunct{\mcitedefaultmidpunct}
{\mcitedefaultendpunct}{\mcitedefaultseppunct}\relax
\EndOfBibitem
\bibitem[Altman and Bayer(2005)]{altman2005human}
J.~Altman and S.~A. Bayer, \emph{The human brain during the second trimester},
  CRC Press, 2005\relax
\mciteBstWouldAddEndPuncttrue
\mciteSetBstMidEndSepPunct{\mcitedefaultmidpunct}
{\mcitedefaultendpunct}{\mcitedefaultseppunct}\relax
\EndOfBibitem
\bibitem[Roussos \emph{et~al.}(2011)Roussos, Condeelis, and
  Patsialou]{Roussos:2011aa}
E.~T. Roussos, J.~S. Condeelis and A.~Patsialou, \emph{Nat. Rev. Cancer}, 2011,
  \textbf{11}, 573 EP --\relax
\mciteBstWouldAddEndPuncttrue
\mciteSetBstMidEndSepPunct{\mcitedefaultmidpunct}
{\mcitedefaultendpunct}{\mcitedefaultseppunct}\relax
\EndOfBibitem
\bibitem[Yin \emph{et~al.}(2009)Yin, Bar-Kochba, and Chen]{Yin2009}
J.~Yin, E.~Bar-Kochba and X.~Chen, \emph{Soft Matter}, 2009, \textbf{5},
  3469--3474\relax
\mciteBstWouldAddEndPuncttrue
\mciteSetBstMidEndSepPunct{\mcitedefaultmidpunct}
{\mcitedefaultendpunct}{\mcitedefaultseppunct}\relax
\EndOfBibitem
\bibitem[Chen and Yin(2010)]{Chen2010}
X.~Chen and J.~Yin, \emph{Soft Matter}, 2010, \textbf{6}, 5667--5680\relax
\mciteBstWouldAddEndPuncttrue
\mciteSetBstMidEndSepPunct{\mcitedefaultmidpunct}
{\mcitedefaultendpunct}{\mcitedefaultseppunct}\relax
\EndOfBibitem
\bibitem[Chen(2012)]{Chen2013}
X.~Chen, \emph{Mechanical Self-Assembly: Science and Applications}, Springer
  Science \& Business Media, 2012\relax
\mciteBstWouldAddEndPuncttrue
\mciteSetBstMidEndSepPunct{\mcitedefaultmidpunct}
{\mcitedefaultendpunct}{\mcitedefaultseppunct}\relax
\EndOfBibitem
\bibitem[Wilcox \emph{et~al.}(2012)Wilcox, Mu{\~n}oz-Navas, and
  Sung]{wilcox2012}
C.~M. Wilcox, M.~Mu{\~n}oz-Navas and J.~J. Sung, \emph{Atlas of Clinical
  Gastrointestinal Endoscopy E-Book: Expert Consult-Online and Print}, Elsevier
  Health Sciences, 2012\relax
\mciteBstWouldAddEndPuncttrue
\mciteSetBstMidEndSepPunct{\mcitedefaultmidpunct}
{\mcitedefaultendpunct}{\mcitedefaultseppunct}\relax
\EndOfBibitem
\bibitem[Zurlo and Truskinovsky(2017)]{zurlo2017}
G.~Zurlo and L.~Truskinovsky, \emph{Phys. Rev. Lett.}, 2017, \textbf{119},
  048001\relax
\mciteBstWouldAddEndPuncttrue
\mciteSetBstMidEndSepPunct{\mcitedefaultmidpunct}
{\mcitedefaultendpunct}{\mcitedefaultseppunct}\relax
\EndOfBibitem
\bibitem[Du \emph{et~al.}(2018)Du, L{\"u}, Chen, and Destrade]{Du2018}
Y.~Du, C.~L{\"u}, W.~Chen and M.~Destrade, \emph{J. Mech. Phys. Solids}, 2018,
  \textbf{118}, 133--151\relax
\mciteBstWouldAddEndPuncttrue
\mciteSetBstMidEndSepPunct{\mcitedefaultmidpunct}
{\mcitedefaultendpunct}{\mcitedefaultseppunct}\relax
\EndOfBibitem
\bibitem[Holzapfel \emph{et~al.}(2007)Holzapfel, Sommer, Auer, Regitnig, and
  Ogden]{Holzapfel2007}
G.~A. Holzapfel, G.~Sommer, M.~Auer, P.~Regitnig and R.~W. Ogden, \emph{Ann.
  Biomed. Eng.}, 2007, \textbf{35}, 530--545\relax
\mciteBstWouldAddEndPuncttrue
\mciteSetBstMidEndSepPunct{\mcitedefaultmidpunct}
{\mcitedefaultendpunct}{\mcitedefaultseppunct}\relax
\EndOfBibitem
\bibitem[Holzapfel and Ogden(2010)]{Holzapfel2010}
G.~A. Holzapfel and R.~W. Ogden, \emph{J. Royal Soc. Interface}, 2010,
  \textbf{7}, 787--799\relax
\mciteBstWouldAddEndPuncttrue
\mciteSetBstMidEndSepPunct{\mcitedefaultmidpunct}
{\mcitedefaultendpunct}{\mcitedefaultseppunct}\relax
\EndOfBibitem
\bibitem[Omens and Fung(1990)]{Omens1990}
J.~H. Omens and Y.-C. Fung, \emph{Circ. Res.}, 1990, \textbf{66}, 37--45\relax
\mciteBstWouldAddEndPuncttrue
\mciteSetBstMidEndSepPunct{\mcitedefaultmidpunct}
{\mcitedefaultendpunct}{\mcitedefaultseppunct}\relax
\EndOfBibitem
\bibitem[Savin \emph{et~al.}(2011)Savin, Kurpios, Shyer, Florescu, Liang,
  Mahadevan, and Tabin]{Savin2011}
T.~Savin, N.~A. Kurpios, A.~E. Shyer, P.~Florescu, H.~Liang, L.~Mahadevan and
  C.~J. Tabin, \emph{Nature}, 2011, \textbf{476}, 57--62\relax
\mciteBstWouldAddEndPuncttrue
\mciteSetBstMidEndSepPunct{\mcitedefaultmidpunct}
{\mcitedefaultendpunct}{\mcitedefaultseppunct}\relax
\EndOfBibitem
\bibitem[Stylianopoulos \emph{et~al.}(2012)Stylianopoulos, Martin, Chauhan,
  Jain, Diop-Frimpong, Bardeesy, Smith, Ferrone, Hornicek, Boucher, Munn, and
  Jain]{Stylianopoulos2012}
T.~Stylianopoulos, J.~D. Martin, V.~P. Chauhan, S.~R. Jain, B.~Diop-Frimpong,
  N.~Bardeesy, B.~L. Smith, C.~R. Ferrone, F.~J. Hornicek, Y.~Boucher, L.~L.
  Munn and R.~K. Jain, \emph{Proc. Natl. Acad. Sci. U.S.A.}, 2012,
  \textbf{109}, 15101--15108\relax
\mciteBstWouldAddEndPuncttrue
\mciteSetBstMidEndSepPunct{\mcitedefaultmidpunct}
{\mcitedefaultendpunct}{\mcitedefaultseppunct}\relax
\EndOfBibitem
\bibitem[Fernandez-Sanchez \emph{et~al.}(2015)Fernandez-Sanchez, Barbier,
  Whitehead, Bealle, Michel, Latorre-Ossa, Rey, Fouassier, Claperon, Brulle,
  Girard, Servant, Rio-Frio, Marie, Lesieur, Housset, Gennisson, Tanter,
  Menager, Fre, Robine, and Farge]{Fernandez-Sanchez2015}
M.~E. Fernandez-Sanchez, S.~Barbier, J.~Whitehead, G.~Bealle, A.~Michel,
  H.~Latorre-Ossa, C.~Rey, L.~Fouassier, A.~Claperon, L.~Brulle, E.~Girard,
  N.~Servant, T.~Rio-Frio, H.~Marie, S.~Lesieur, C.~Housset, J.~L. Gennisson,
  M.~Tanter, C.~Menager, S.~Fre, S.~Robine and E.~Farge, \emph{Nature}, 2015,
  \textbf{523}, 92--95\relax
\mciteBstWouldAddEndPuncttrue
\mciteSetBstMidEndSepPunct{\mcitedefaultmidpunct}
{\mcitedefaultendpunct}{\mcitedefaultseppunct}\relax
\EndOfBibitem
\bibitem[Pan \emph{et~al.}(2016)Pan, Heemskerk, Ibar, Shraiman, and
  Irvine]{Pan2016}
Y.~Pan, I.~Heemskerk, C.~Ibar, B.~I. Shraiman and K.~D. Irvine, \emph{Proc.
  Natl. Acad. Sci. U.S.A.}, 2016,  201615012\relax
\mciteBstWouldAddEndPuncttrue
\mciteSetBstMidEndSepPunct{\mcitedefaultmidpunct}
{\mcitedefaultendpunct}{\mcitedefaultseppunct}\relax
\EndOfBibitem
\bibitem[Padera \emph{et~al.}(2004)Padera, Stoll, Tooredman, Capen, Tomaso, and
  Jain]{Padera2004}
T.~P. Padera, B.~R. Stoll, J.~B. Tooredman, D.~Capen, E.~D. Tomaso and R.~K.
  Jain, \emph{Nature}, 2004, \textbf{427}, 695\relax
\mciteBstWouldAddEndPuncttrue
\mciteSetBstMidEndSepPunct{\mcitedefaultmidpunct}
{\mcitedefaultendpunct}{\mcitedefaultseppunct}\relax
\EndOfBibitem
\bibitem[Helmlinger \emph{et~al.}(1997)Helmlinger, Netti, Lichtenbeld, Melder,
  and Jain]{Helmlinger1997}
G.~Helmlinger, P.~A. Netti, H.~C. Lichtenbeld, R.~J. Melder and R.~K. Jain,
  \emph{Nat. Biotechnol.}, 1997, \textbf{15}, 778\relax
\mciteBstWouldAddEndPuncttrue
\mciteSetBstMidEndSepPunct{\mcitedefaultmidpunct}
{\mcitedefaultendpunct}{\mcitedefaultseppunct}\relax
\EndOfBibitem
\bibitem[Cao \emph{et~al.}(2012)Cao, Li, and Feng]{Cao2012}
Y.-P. Cao, B.~Li and X.-Q. Feng, \emph{Soft Matter}, 2012, \textbf{8},
  556--562\relax
\mciteBstWouldAddEndPuncttrue
\mciteSetBstMidEndSepPunct{\mcitedefaultmidpunct}
{\mcitedefaultendpunct}{\mcitedefaultseppunct}\relax
\EndOfBibitem
\bibitem[Li \emph{et~al.}(2011)Li, Cao, Feng, and Gao]{Li2011}
B.~Li, Y.-P. Cao, X.-Q. Feng and H.~Gao, \emph{J. Mech. Phys. Solids}, 2011,
  \textbf{59}, 758--774\relax
\mciteBstWouldAddEndPuncttrue
\mciteSetBstMidEndSepPunct{\mcitedefaultmidpunct}
{\mcitedefaultendpunct}{\mcitedefaultseppunct}\relax
\EndOfBibitem
\bibitem[L{\"u} and Du(2016)]{Lu2016}
C.~L{\"u} and Y.~Du, \emph{Int. J. Appl. Mech.}, 2016, \textbf{8},
  1640010\relax
\mciteBstWouldAddEndPuncttrue
\mciteSetBstMidEndSepPunct{\mcitedefaultmidpunct}
{\mcitedefaultendpunct}{\mcitedefaultseppunct}\relax
\EndOfBibitem
\bibitem[Du and L{\"u}(2017)]{Du2017}
Y.~Du and C.~L{\"u}, \emph{Theor. Appl. Mech. Lett.}, 2017, \textbf{7},
  117--120\relax
\mciteBstWouldAddEndPuncttrue
\mciteSetBstMidEndSepPunct{\mcitedefaultmidpunct}
{\mcitedefaultendpunct}{\mcitedefaultseppunct}\relax
\EndOfBibitem
\bibitem[Rodriguez \emph{et~al.}(1994)Rodriguez, Hoger, and
  Mcculloch]{Rodriguez1994}
E.~K. Rodriguez, A.~Hoger and A.~D. Mcculloch, \emph{J. Biomech.}, 1994,
  \textbf{27}, 455--467\relax
\mciteBstWouldAddEndPuncttrue
\mciteSetBstMidEndSepPunct{\mcitedefaultmidpunct}
{\mcitedefaultendpunct}{\mcitedefaultseppunct}\relax
\EndOfBibitem
\bibitem[Goriely and {Ben Amar}(2005)]{Goriely2005}
A.~Goriely and M.~{Ben Amar}, \emph{Phys. Rev. Lett.}, 2005, \textbf{94},
  198103\relax
\mciteBstWouldAddEndPuncttrue
\mciteSetBstMidEndSepPunct{\mcitedefaultmidpunct}
{\mcitedefaultendpunct}{\mcitedefaultseppunct}\relax
\EndOfBibitem
\bibitem[Ciarletta(2013)]{Ciarletta2013}
P.~Ciarletta, \emph{Phys. Rev. Lett.}, 2013, \textbf{110}, 158102\relax
\mciteBstWouldAddEndPuncttrue
\mciteSetBstMidEndSepPunct{\mcitedefaultmidpunct}
{\mcitedefaultendpunct}{\mcitedefaultseppunct}\relax
\EndOfBibitem
\bibitem[Du \emph{et~al.}(2019)Du, L{\"u}, Chen, and Destrade]{Du2019}
Y.~Du, C.~L{\"u}, W.~Chen and M.~Destrade, \emph{Sc. Reports}, 2019,
  \textbf{9}, 8232\relax
\mciteBstWouldAddEndPuncttrue
\mciteSetBstMidEndSepPunct{\mcitedefaultmidpunct}
{\mcitedefaultendpunct}{\mcitedefaultseppunct}\relax
\EndOfBibitem
\bibitem[Sadik and Yavari(2015)]{sadik2015}
S.~Sadik and A.~Yavari, \emph{Math. Mech. Solids}, 2015, \textbf{22},
  771--772\relax
\mciteBstWouldAddEndPuncttrue
\mciteSetBstMidEndSepPunct{\mcitedefaultmidpunct}
{\mcitedefaultendpunct}{\mcitedefaultseppunct}\relax
\EndOfBibitem
\bibitem[Ciarletta \emph{et~al.}(2014)Ciarletta, Balbi, and
  Kuhl]{Ciarletta2014}
P.~Ciarletta, V.~Balbi and E.~Kuhl, \emph{Phys. Rev. Lett.}, 2014,
  \textbf{113}, 248101\relax
\mciteBstWouldAddEndPuncttrue
\mciteSetBstMidEndSepPunct{\mcitedefaultmidpunct}
{\mcitedefaultendpunct}{\mcitedefaultseppunct}\relax
\EndOfBibitem
\bibitem[Pedersen(2006)]{pedersen2006shrink}
P.~Pedersen, \emph{Computational Mechanics}, 2006, \textbf{37}, 121--130\relax
\mciteBstWouldAddEndPuncttrue
\mciteSetBstMidEndSepPunct{\mcitedefaultmidpunct}
{\mcitedefaultendpunct}{\mcitedefaultseppunct}\relax
\EndOfBibitem
\bibitem[Ogden(1997)]{ogden1997non}
R.~W. Ogden, \emph{Non-linear elastic deformations}, Courier Corporation,
  1997\relax
\mciteBstWouldAddEndPuncttrue
\mciteSetBstMidEndSepPunct{\mcitedefaultmidpunct}
{\mcitedefaultendpunct}{\mcitedefaultseppunct}\relax
\EndOfBibitem
\bibitem[Biryukov(1985)]{Biryukov1985350}
S.~Biryukov, \emph{Sov. Phys. Acoust.}, 1985, \textbf{31}, 350--354\relax
\mciteBstWouldAddEndPuncttrue
\mciteSetBstMidEndSepPunct{\mcitedefaultmidpunct}
{\mcitedefaultendpunct}{\mcitedefaultseppunct}\relax
\EndOfBibitem
\bibitem[Destrade \emph{et~al.}(2010)Destrade, Murphy, and
  Ogden]{DESTRADE20101212}
M.~Destrade, J.~Murphy and R.~Ogden, \emph{Int. J. Eng. Sci.}, 2010,
  \textbf{48}, 1212 -- 1224\relax
\mciteBstWouldAddEndPuncttrue
\mciteSetBstMidEndSepPunct{\mcitedefaultmidpunct}
{\mcitedefaultendpunct}{\mcitedefaultseppunct}\relax
\EndOfBibitem
\bibitem[Destrade \emph{et~al.}(2009)Destrade, {N{\'\i} Annaidh}, and
  Coman]{DESTRADE20094322}
M.~Destrade, A.~{N{\'\i} Annaidh} and C.~D. Coman, \emph{Int. J. Solids
  Struct.}, 2009, \textbf{46}, 4322 -- 4330\relax
\mciteBstWouldAddEndPuncttrue
\mciteSetBstMidEndSepPunct{\mcitedefaultmidpunct}
{\mcitedefaultendpunct}{\mcitedefaultseppunct}\relax
\EndOfBibitem
\bibitem[Su \emph{et~al.}(2019)Su, Wu, Chen, and Destrade]{SU2019}
Y.~Su, B.~Wu, W.~Chen and M.~Destrade, \emph{J. Mech. Phys. Solids}, 2019\relax
\mciteBstWouldAddEndPuncttrue
\mciteSetBstMidEndSepPunct{\mcitedefaultmidpunct}
{\mcitedefaultendpunct}{\mcitedefaultseppunct}\relax
\EndOfBibitem
\bibitem[Ciarletta \emph{et~al.}(2016)Ciarletta, Destrade, Gower, and
  Taffetani]{Ciarletta2016}
P.~Ciarletta, M.~Destrade, A.~L. Gower and M.~Taffetani, \emph{J. Mech. Phys.
  Solids}, 2016, \textbf{90}, 242--253\relax
\mciteBstWouldAddEndPuncttrue
\mciteSetBstMidEndSepPunct{\mcitedefaultmidpunct}
{\mcitedefaultendpunct}{\mcitedefaultseppunct}\relax
\EndOfBibitem
\bibitem[Cao and Hutchinson(2011)]{cao2011wrinkles}
Y.~Cao and J.~W. Hutchinson, \emph{Proc. Roy. Soc. A}, 2011, \textbf{468},
  94--115\relax
\mciteBstWouldAddEndPuncttrue
\mciteSetBstMidEndSepPunct{\mcitedefaultmidpunct}
{\mcitedefaultendpunct}{\mcitedefaultseppunct}\relax
\EndOfBibitem
\end{mcitethebibliography}
\end{document}